\documentclass[10pt, a4paper]{article}
\usepackage[margin=1in]{geometry}
\usepackage{amsmath,amssymb,amsthm}
\usepackage{array}
\usepackage[shortlabels]{enumitem}
\usepackage{graphicx}
\usepackage{subcaption}
\usepackage{thm-restate}

\usepackage[ruled, noend, noline]{algorithm2e}

\usepackage{authblk}

\newtheorem{theorem}{Theorem}
\newtheorem{lemma}{Lemma}

\newtheorem{remark}{Remark}

\newcommand{\floor}[1]{\left\lfloor #1 \right\rfloor}
\newcommand{\ceil}[1]{\left\lceil #1 \right\rceil}

\newcommand{\istar}{i^\star}

\newcommand{\Jstar}{J^\star}

\newcommand{\vhat}{\widehat{v}}
\newcommand{\nB}{J}

\newcommand{\Tpred}{\widehat{T}}
\newcommand{\Jpred}{\widehat{J}}

\newcommand{\con}{\chi}
\newcommand{\rob}{\rho}

\DeclareMathOperator{\bopt}{B}

\DeclareMathOperator{\opt}{OPT}

\DeclareMathOperator{\optval}{opt}

\title{On Optimal Consistency-Robustness Trade-Off for Learning-Augmented Multi-Option Ski Rental}
\author[]{Yongho Shin}
\author[]{Changyeol Lee}
\author[]{Hyung-Chan An\thanks{Corresponding author: \texttt{hyung-chan.an@yonsei.ac.kr}}}
\affil[]{Department of Computer Science, Yonsei University, Seoul, South Korea}
\date{}

\begin{document}

\maketitle
\begin{abstract}
The \emph{learning-augmented multi-option ski rental problem} generalizes the classical ski rental problem in two ways: the algorithm is provided with a \emph{prediction} on the number of days we can ski, and the ski rental options now come with a variety of rental periods and prices to choose from, unlike the classical two-option setting. Subsequent to the initial study of the multi-option ski rental problem (without learning augmentation) due to Zhang, Poon, and Xu, significant progress has been made for this problem recently in particular. The problem is very well understood when we relinquish one of the two generalizations---for the learning-augmented \emph{classical} ski rental problem, algorithms giving best-possible trade-off between consistency and robustness exist; for the multi-option ski rental problem without learning augmentation, deterministic/randomized algorithms giving the best-possible competitiveness have been found. However, in presence of both generalizations, there remained a huge gap between the algorithmic and impossibility results. In fact, for randomized algorithms, we did not have any nontrivial lower bounds on the consistency-robustness trade-off before.

This paper bridges this gap for both deterministic and randomized algorithms. For deterministic algorithms, we present a best-possible algorithm that completely matches the known lower bound. For randomized algorithms, we show the first nontrivial lower bound on the consistency-robustness trade-off, and also present an improved randomized algorithm. Our algorithm matches our lower bound on robustness within a factor of $e/2$ when the consistency is at most $1.086$.
\end{abstract}

\section{Introduction} \label{sec:intro}
The \emph{learning-augmented multi-option ski rental problem} is a generalization of  classical ski rental. In this problem, we are required to choose from multiple ski rental options so that we have a pair of skis available as long as the ski resort is open. The number of days for which the resort will be open is not known in advance, making this problem an online optimization, yet a \emph{prediction} on the number of days is provided to the algorithm. With the help of this prediction, the algorithm has to ensure that a pair of skis is available by choosing from a multiple number of rental options that come with a variety of rental periods and costs. Naturally, the objective is to minimize the total cost paid.

This problem generalizes the classical ski rental problem in two ways. Firstly, the algorithm is provided with a prediction on the number of days, which does not exist in the classical setting. This prediction is usually obtained via machine learning (ML). As such, while the prediction may be empirically accurate, there is no guarantee whatsoever on the quality of this prediction. The challenge is therefore in obtaining an algorithm that can effectively exploit the prediction when it is accurate while at the same time guaranteeing a certain ``minimum'' level of performance even when the prediction is bad. Secondly, there can be more than two ski rental options in this problem. In the classical two-option problem, skis can be either rented for a single day or purchased for good. This problem lifts this restriction and allows rental options that rent a pair of skis for a finite number of days at a certain cost. These rental periods and costs are given as part of the input.

This problem is very well-understood when we relinquish one of these two generalizations. For the learning-augmented classical (two-option) ski rental problem, Kumar, Purohit, and Svitkina~\cite{kumar2018improving} gave a best-possible deterministic algorithm for this problem. Learning-augmented algorithms are often evaluated by analyzing their \emph{consistency} and \emph{robustness}~\cite{lykouris2021competitive, kumar2018improving}: we say an algorithm is $\chi$-consistent and $\rho$-robust if it produces a solution whose cost is within a factor of $\chi$ when the given prediction is accurate and within a factor of $\rho$ no matter how bad the prediction is. Kumar et al.'s deterministic algorithm is $(1+\lambda)$-consistent and $(1+1/\lambda)$-robust, where $\lambda \in (0,1)$ is a parameter taken by the algorithm; Angelopoulos, D\"urr, Jin, Kamali, and Renault~\cite{angelopoulos2020online} showed that this is a best-possible for a deterministic algorithm. For randomized algorithms, Kumar et al.~\cite{kumar2018improving} gave an algorithm that was later shown to be asymptotically best possible due to Wei and Zhang~\cite{wei2020optimal} and Bamas, Maggiori, and Svensson~\cite{bamas2020primal}.

For the multi-option ski rental problem without learning augmentation, Zhang, Poon, and Xu~\cite{zhang2011ski} gave a deterministic 4-competitive algorithm, along with a matching lower bound on the competitiveness of deterministic algorithms. Their algorithm, however, relied on a mild assumption that the per-day costs of the options are monotone with respect to rental period, which was later lifted by a general 4-competitive algorithm of Anand, Ge, Kumar, and Panigrahi~\cite{anand2021regression}. For randomized algorithms, Shin, Lee, Lee, and An~\cite{shin2023improved} gave a best-possible $e$-competitive algorithm; they also showed a matching lower bound on the competitive ratio.

In presence of both generalizations, however, the learning-augmented multi-option ski rental problem had a huge gap between the known algorithmic results and the impossibility results, despite the significant recent progress in this problem~\cite{anand2021regression, shin2023improved}. On the algorithmic side, Anand et al.~\cite{anand2021regression} gave the first deterministic algorithm that is $(1 + \varepsilon)$-consistent and $(5 + 5/\varepsilon)$-robust for $\varepsilon > 0$, which was improved by Shin et al.'s $\max(1 + 2\lambda, 4\lambda)$-consistent $(2 + 2/\lambda)$-robust deterministic algorithm~\cite{shin2023improved} for $\lambda \in [0, 1]$; they also gave a randomized $\chi(\lambda)$-consistent $e^\lambda/\lambda$-robust algorithm for  $\lambda \in [0, 1]$, where
$\chi(\lambda) := \{\begin{subarray}{l}1 + \lambda,\\ (e + 1) \lambda - \ln \lambda - 1,~\end{subarray} \begin{subarray}{l} \textrm{if } \lambda < 1/e, \\\textrm{otherwise}.\end{subarray}$
\linebreak On the lower bounds side, however, the best bound known for deterministic algorithms was that, for any $\lambda \in (0, 1)$, a $(1 + \lambda)$-consistent algorithm cannot be better than $(2 + \lambda + 1/\lambda)$-robust~\cite{shin2023improved}, leaving a huge gap between the best deterministic algorithm known. Our understanding was even poorer for randomized algorithms: no nontrivial lower bounds on the consistency-robustness trade-off of randomized algorithms were previously known.

In this paper, we bridge this gap in our understanding of the learning-augmented multi-option ski rental problem, for both deterministic and randomized algorithms. 
Following are the main results  presented in this paper.
\begin{itemize}
\item We present a  deterministic $\frac{1}{(1-\lambda)}$-consistent $\frac{1}{\lambda(1-\lambda)}$-robust algorithm for the problem, parameterized by $\lambda \in [0, 1/2]$.\footnote{We note that, when $\lambda=1/2$, the algorithm becomes $4$-robust. Since this matches the lower bound on the competitiveness without learning augmentation, there is no hope that we can further improve the robustness of the algorithm and therefore we do not consider any higher value of $\lambda$.}
This consistency-robustness trade-off matches the lower bound of Shin et al.~\cite{shin2023improved}, showing that our algorithm is a best-possible deterministic algorithm.
Interestingly, despite being a best-possible algorithm, both the algorithm and its analysis are significantly simpler than previous algorithms~\cite{shin2023improved}.
\item We present a randomized $\chi(\delta, s)$-consistent $\rho(\delta, s)$-robust algorithm parameterized by $\delta \geq e$ and $s \geq 0$, where
\[
\chi(\delta, s) := \begin{cases}
1 + \frac{\delta^{-s}}{\ln \delta}, & s > 1, \\
\frac{\delta + 1}{\ln \delta} \delta^{-s} + s - \frac{1}{\ln \delta}, & 0 \leq s \leq 1,
\end{cases}
\text{ and }
\rho(\delta, s) := \frac{\delta}{e \ln \delta} \cdot \frac{e^{\delta^{-s}}}{\delta^{-s}}.
\]This improves upon the best trade-off attained by previous randomized algorithms~\cite{shin2023improved}.
\item We provide the first nontrivial lower bound for randomized learning-augmented algorithms. We prove that no $(1 + \lambda)$-consistent algorithm can have a robustness better than $\frac{(1 + \lambda)^2}{2\lambda}$ for all $\lambda \in (0, 1)$. We note that this bound is within a constant factor of our randomized algorithm's performance when $\lambda$ is small, i.e., when the prediction is relatively well trusted.
\end{itemize}

\begin{figure}
\begin{subfigure}{0.5\textwidth}
\centering
\includegraphics[width=\textwidth]{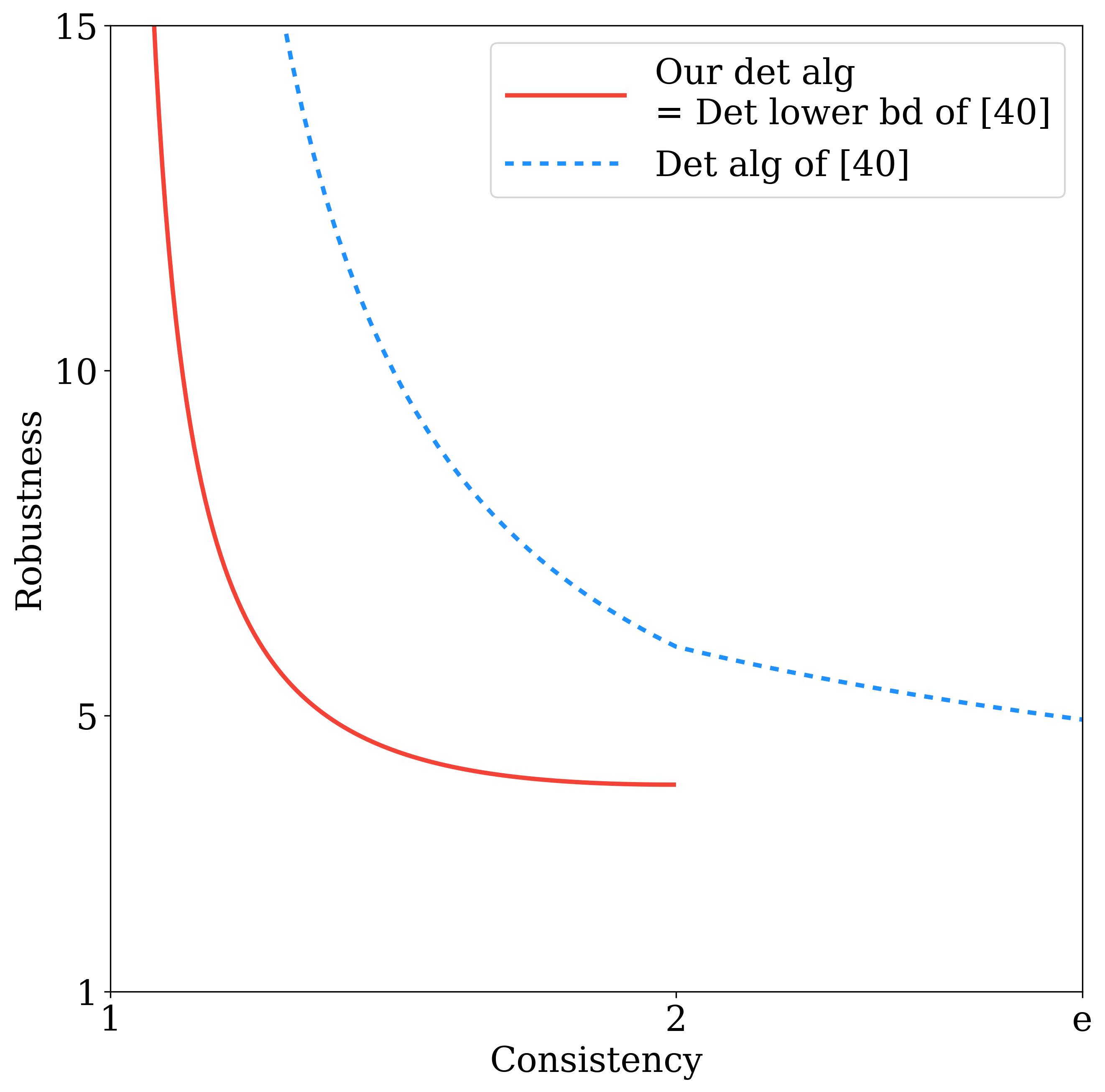}
%\caption{(a) hello}
\end{subfigure}
\begin{subfigure}{0.5\textwidth}
\centering
\includegraphics[width=\textwidth]{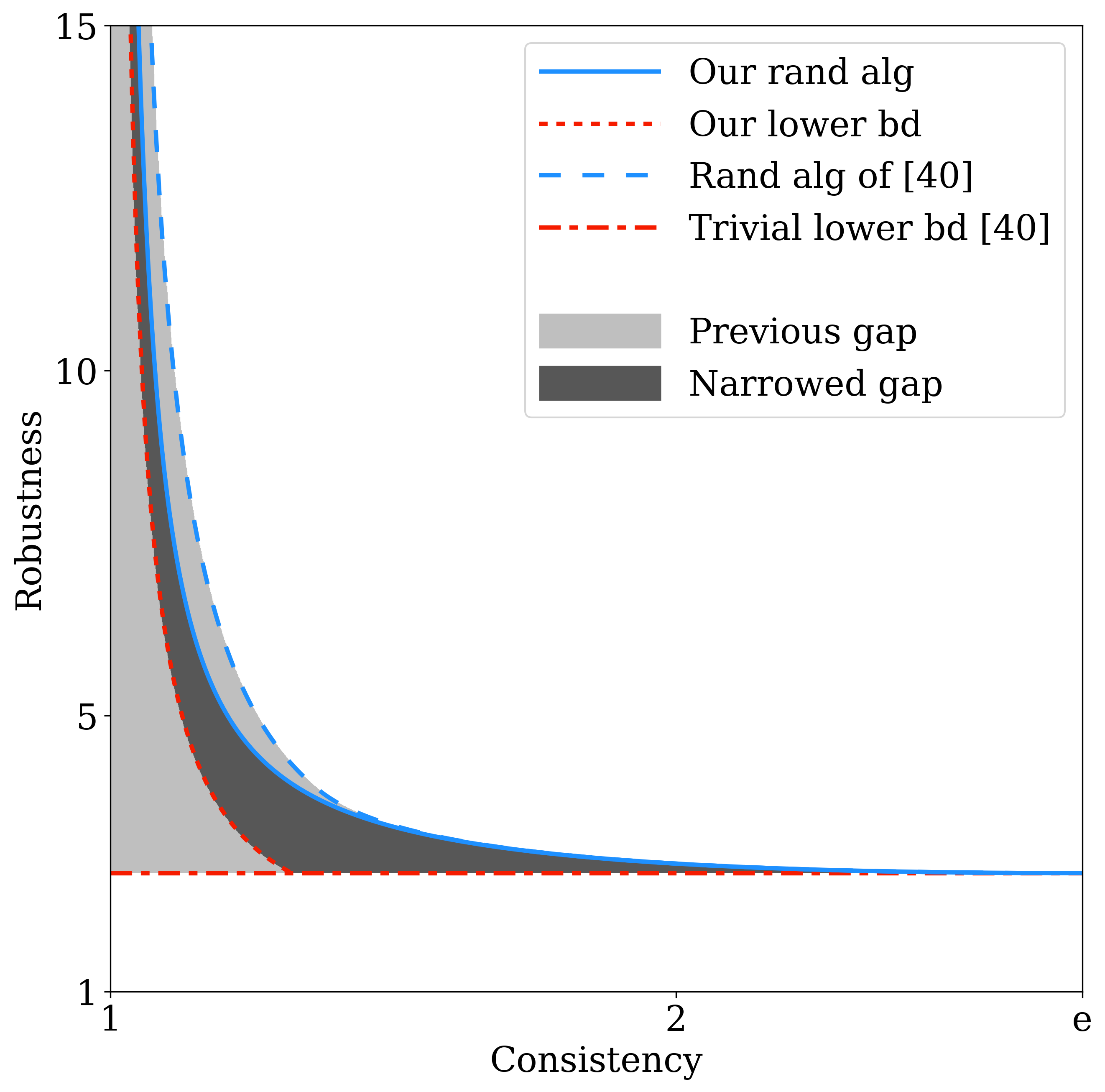}
%\caption{(a) hello}
\end{subfigure}
\caption{Overview of our results. (Left) The red solid line depicts the trade-off of our best-possible deterministic algorithm, whereas the blue dotted line represent the trade-off of Shin et al.'s deterministic algorithm~\cite{shin2023improved}. (Right) The trade-off of our randomized algorithm is drawn as the blue solid line and that of Shin et al.'s randomized algorithm is shown as the blue dashed line. The red dotted line depicts our lower bound on the trade-off. The red dash-dotted line indicates the trivial lower bound of $e$~\cite{shin2023improved}. The dark gray region depicts the new gap between the algorithmic and impossibility bound, narrowing the previous gap marked as the light+dark gray region.}
\label{fig:con-rob}
\end{figure}

Figure~\ref{fig:con-rob} summarizes our results. The graph on the left shows the gap that existed between the best algorithm and the best lower bound known for deterministic algorithms. Our new algorithm (red solid line) matches the previously known lower bound. For randomized algorithms, no nontrivial lower bounds were previously known, and the gap between algorithms and lower bounds was rather wide (light+dark gray region). We present an improved randomized algorithm (blue solid line) and the first nontrivial lower bound (red dotted line) to significantly narrow this gap, shown as the dark gray region.

Section~\ref{sec:det} presents our best-possible deterministic algorithm. Our improved randomized algorithm is presented in Section~\ref{sec:rand}. Section~\ref{sec:lb} then presents the first lower bound on the consistency-robustness trade-off of randomized algorithms.

\paragraph*{Related Work}
Since the seminal work of Lykouris and Vassilvitskii~\cite{lykouris2021competitive}, a tremendous amount of research on learning-augmented algorithms has surged. This algorithmic paradigm gives a sweet breakthrough for online optimization in particular; many online optimization problems suffer from pessimistic guarantees in the worst case since the full information of the input is not given while an irrevocable decision should be made for each timestep. However, we can improve the performance guarantee when we are given a prediction on the future data. To name a few examples of successful augmentation of prediction to online optimization problems, it has been studied for caching/paging~\cite{lykouris2021competitive, rohatgi2020near, antoniadis2023online, wei2020better, im2022parsimonious}, weighted paging~\cite{jiang2022onlinealgorithms, bansal2022learning},
ski rental~\cite{gollapudi2019online, anand2020customizing},
scheduling problems~\cite{kumar2018improving, mitzenmacher2020scheduling, wei2020optimal, im2021non, azar2021flow, lindermayr2022permutation},
load balancing~\cite{lattanzi2020online, lavastida2021learnable},
energy minimization~\cite{bamas2020learning},
matching problems~\cite{antoniadis2020secretary, lavastida2021using, jin2022online},
network design problems~\cite{xu2022learning, erlebach2022learning},
optimization problems in metric spaces~\cite{antoniadis2023online, almanza2021online, lindermayr2022double, jiang2022onlinefacility, azar2022online},
and convex function chasing~\cite{christianson2022chasing}.
Learning-augmented algorithms have also been used to improve an algorithm's running time~\cite{kraska2018case, dinitz2021faster}. We refer interested readers to the survey of Mitzenmacher and Vassilvitskii~\cite{mitzenmacher2022algorithms} for a gentle introduction to learning-augmented algorithms.

The ski rental problem is a canonical online optimization problem, and has been intensively studied. For the classical two-option problem, Karlin, Manasse, Rudolph, and Sleator~\cite{karlin1986competitive} gave a deterministic $2$-competitive algorithm and Karlin, Manasse, McGeoch, and Owicki~\cite{karlin1990competitive} gave a randomized $e/(e-1)$-competitive algorithm. Both algorithms are best possible. Ski rental problems have been widely studied under various settings including, for example, multi-shop ski rental~\cite{ai2014multi, wang2020online}, snoopy caching~\cite{karlin1986competitive, karlin1990competitive}, dynamic TCP acknowledgment~\cite{karlin2001dynamic, banerjee2020improving}, the parking permit problem~\cite{meyerson2005parking}, the Bahncard problem~\cite{fleischer2001bahncard}, and applications to online cloud file systems~\cite{khanafer2013constrained}.

\section{Preliminaries}
In the \emph{learning-augmented multi-option ski rental problem}, we are given as input a set of options for renting skis and a prediction $\Tpred$ on the number of days for which the ski resort will be open.
For each option~$i$, we are given $c_i \in \mathbb{Q}_{>0}$ and $d_i \in \mathbb{Z}_{>0} \cup \{\infty\}$: when we rent option~$i$, we pay the cost of $c_i$ and can use skis for $d_i$ days from the day of renting. Renting for $\infty$ days corresponds to buying. Without loss of generality, let us assume that $c_i \geq 1$ for every option~$i$; otherwise, we may multiply all $c_i$'s by a sufficiently large number.

\newpage

On each day, we learn whether the ski resort is open for that day; if the resort is open, but we have no skis available for the day, we need to choose one of the rental options and pay for it. Let $T$ be the number of days the resort is open; the objective is to have skis available for the entire $T$ days at the minimum cost.

For every $t \in \mathbb{Z}_{>0}$, let $\opt(t)$ denote an optimal solution (i.e., a minimum-cost sequence of rental options) that covers $t$ days at minimum cost. Let $\optval(t)$ denote the total cost incurred by $\opt(t)$. Note that, by a standard dynamic programming, we can easily compute $\opt(t)$ and $\optval(t)$ for any $t$. We therefore assume in what follows that, for all $t$, $\opt(t)$ and $\optval(t)$ are readily available whenever we want to use it. Note that $\optval(t) \geq 1$ for all $t$, due to our assumption that $c_i \geq 1$ for every option~$i$.

A standard way of measuring the performance of learning-augmented algorithms is the consistency-robustness trade-off analysis~\cite{lykouris2021competitive, kumar2018improving}. We say that a learning-augmented algorithm for this problem is \emph{$\chi$-consistent}  if the (expected) cost incurred by the algorithm is at most $\chi \cdot\optval(T)$ when the prediction is accurate (i.e., $\Tpred=T$). On the other hand, we say that an algorithm is \emph{$\rho$-robust}  if the (expected) cost of the algorithm's solution does not exceed $\rho \cdot\optval(T)$ no matter how (in)accurate the prediction is (i.e., for any $\Tpred$).

We introduce some definitions before we present our algorithms. Given two solutions $S_1$ and $S_2$, we say that we \emph{append} $S_1$ to $S_2$ when we concatenate $S_1$ after $S_2$. That is, the concatenated solution is to pay for each option in $S_2$ and then for each option in $S_1$. (If an option appears a multiple number of times, we choose and pay for it each time it appears.) For all $v \geq 1$, let $\bopt(v)$ be a solution covering the most number of days among those whose cost does not exceed $v$: i.e., $\bopt(v) := \opt(t^{\star})$ where $t^{\star} := \max \{t \in \mathbb{Z}_{>0} \cup \{\infty\} \mid \optval(t) \leq v\}$. We set $\bopt(v)$ as an empty solution if $\{t \in \mathbb{Z}_{>0} \cup \{\infty\} \mid \optval(t) \leq v\} = \emptyset$.

Finally, for simplicity of presentation, we will describe our algorithms as if they never terminate and keep choosing rental options; however, this is to be interpreted really as an algorithm that gets immediately halted once the solution output by the algorithm so far covers the last day $T$.

\section{Best-Possible Deterministic Algorithm}\label{sec:det}
In this section, we present our deterministic  algorithm for the learning-augmented multi-option ski rental problem.
This algorithm is best possible for a deterministic algorithm; moreover, it admits a much simpler analysis than previous algorithms.

The algorithm takes an input parameter $\lambda \in [0, 1/2]$.
Let us assume that $\lambda > 0$; we will later discuss how to handle $\lambda = 0$. We can assume without loss of generality that $\optval(\Tpred) = (1/\lambda)^k$ for some integer $k$ since, if $ (1/\lambda)^{k - 1} < \optval(\Tpred) < (1/\lambda)^k $, we may multiply the cost of every option by $\frac{(1/\lambda)^k}{\optval(\Tpred)}$.

The algorithm is very simple: the algorithm consists of several \emph{iterations}, and in each iteration~$i$ (for $i = 0, 1, 2, \ldots$), we append $\bopt((1/\lambda)^i)$ to our solution.

\begin{restatable}{theorem}{detTheorem} \label{thm:optdet}
For $\lambda \in (0, 1/2]$, the given algorithm is a deterministic $\frac{1}{1-\lambda}$-consistent $\frac{1}{\lambda(1-\lambda)}$-robust algorithm.
\end{restatable}
\begin{proof}
\emph{Consistency.} Suppose $T=\Tpred$. Observe that the algorithm terminates at iteration~$k$ (or earlier) by the fact that $\optval(\Tpred) = (1/\lambda)^k$ and the definition of $\bopt(\cdot)$. Moreover, in each iteration~$i$, the algorithm incurs the cost of at most $(1/\lambda)^i$ from the definition of $\bopt(\cdot)$. Hence, the total cost incurred by the algorithm is at most
$
\sum_{i = 0}^k (1/\lambda)^i \leq \frac{(1/\lambda)^{k + 1}}{(1/\lambda) - 1},
$
implying that the consistency ratio is at most $\frac{(1/\lambda)}{(1/\lambda) - 1} = \frac{1}{1 - \lambda}$ as desired.

\emph{Robustness.} Suppose that $\optval(T) = (1/\lambda)^{\istar}$ for some $\istar \in \mathbb{R}_{\geq 0}$. Again by the definition of $\bopt(\cdot)$, note that the algorithm terminates at iteration~$\ceil{\istar}$ (or earlier). Therefore, the total cost incurred by the algorithm is at most
$
\sum_{i = 0}^{\ceil{\istar}} (1/\lambda)^i \leq \frac{(1/\lambda)^{\ceil{\istar} + 1}}{(1/\lambda) - 1} \leq \frac{(1/\lambda)^{\istar + 2}}{(1/\lambda) - 1},
$
giving the desired robustness ratio since we have $\frac{(1/\lambda)^2}{(1/\lambda) - 1} = \frac{1}{\lambda (1 - \lambda)}$.
\end{proof}

\begin{remark}
For $\lambda = 0$, we can easily obtain a $1$-consistent $\infty$-robust algorithm: consider an algorithm that appends $\opt(\Tpred)$ at the very beginning of the execution.
\end{remark}

We now show that our deterministic algorithm is best possible. Shin et al.~\cite{shin2023improved} gave the following lower bound for deterministic algorithms.
\begin{theorem}[\cite{shin2023improved}, Theorem 6] \label{thm:detlb}
For all constant $c \in (1, 2)$ and $\varepsilon > 0$, the robustness ratio of any deterministic $c$-consistent algorithm must be greater than $c^2 / (c - 1) - \varepsilon$.
\end{theorem}
Let us substitute $c := 1/(1 - \lambda)$ in Theorem~\ref{thm:optdet}. We can then easily see that
$
\frac{1}{\lambda (1-\lambda)} = \frac{c^2}{c - 1},
$
showing that our algorithm is the best possible.

\section{Improved Randomized Algorithm}\label{sec:rand}
This section is devoted to an improved randomized learning-augmented algorithm for the multi-option ski rental problem. The algorithm takes two parameters $\delta$ and $s$ that adjust the trade-off between the consistency and robustness of the algorithm. We prove the following theorem in this section.
\begin{theorem} \label{thm:rand-param}
For all $\delta \geq e$ and $s \geq 0$, there exists a randomized $\chi(\delta, s)$-consistent $\rho(\delta, s)$-robust algorithm for the multi-option ski rental problem, where
\[
\chi(\delta, s) := \begin{cases}
1 + \frac{\delta^{-s}}{\ln \delta}, & s > 1, \\
\frac{\delta + 1}{\ln \delta} \delta^{-s} + s - \frac{1}{\ln \delta}, & 0 \leq s \leq 1,
\end{cases}
\text{ and }
\rho(\delta, s) := \frac{\delta}{e \ln \delta} \cdot \frac{e^{\delta^{-s}}}{\delta^{-s}}.
\]
\end{theorem}

Similarly to the previous section, let us assume without loss of generality that $\optval(\Tpred) = \delta^k$ for some integer $k \geq s + 2$. Recall that, for all $v \geq 1$, $\bopt(v)$ is a solution that covers the most number of days among those whose cost does not exceed $v$.

\subsection{Our algorithm}
At the beginning, we first sample $\alpha \in [1, \delta)$ from a distribution whose probability density function $f$ is given by
$
f(\alpha) := \frac{1}{\alpha \ln \delta}.
$
Note that $f$  indeed defines a probability distribution.

The algorithm consists of \emph{three} phases. In the first phase, the algorithm runs $k$ iterations named iteration~$i$ for $i = 0, 1, \ldots, k - 1$. In iteration $i$, 
if $\alpha \delta^i < \delta^{k - s}$, we append $\bopt(\alpha \delta^i)$ and continue to the next iteration; if $\alpha \delta^i \geq \delta^{k - s}$, we immediately proceed to the second phase without appending anything. In iteration~$k - 1$, if $\alpha \delta^{k - 1} < \delta^{k - s}$, we append $\bopt(\alpha \delta^{k - 1})$ and proceed to the third phase (skipping the second one); if $\alpha \delta^{k - 1} \geq \delta^{k - s}$, we proceed to the second phase without appending anything.

In the second phase, we append $\opt(\Tpred)$ and proceed to the third phase.

The third phase also consists of iterations. They are named iteration~$i$ for $i = k, k+1, \ldots$, starting from $k$. In each iteration~$i$ of this phase, we append $\bopt(\alpha \delta^i)$.

We have also provided a pseudocode of our algorithm. See Algorithm~\ref{alg:rand}.
\begin{algorithm}
\caption{Our randomized  algorithm}\label{alg:rand}
\SetKwInput{KwData}{Given}
$\triangleright\ $ \textsf{\emph{initialization}}\\\Indp 
sample $\alpha \in [1, \delta)$ from p.d.f. $f(\alpha) := 1 / (\alpha \ln \delta)$\\
\Indm $\triangleright\ $ \textsf{\emph{first phase}}\\\Indp 
\For{$i = 0, 1, \ldots, k - 2$}{
	\If{$\alpha \delta^{i} < \delta^{k - s}$}{append $\bopt(\alpha \delta^{i})$}
	\Else{proceed to the second phase}
}
$i\gets k-1$\\
\If{$\alpha \delta^i < \delta^{k - s}$} {
append $\bopt(\alpha \delta^i)$ \\
proceed to the third phase
}
\Else{proceed to the second phase}
\Indm $\triangleright\ $ \textsf{\emph{second phase}}\\\Indp
append $\opt(\Tpred)$ \\
proceed to the third phase\\
\Indm $\triangleright\ $ \textsf{\emph{third phase}}\\\Indp 
\For{$i = k, k + 1, \ldots$}{
append $\bopt(\alpha \delta^i)$
}
\Indm
\end{algorithm}

\subsection{Analysis}\label{subsec:rand-analysis}
We now prove Theorem~\ref{thm:rand-param}. We begin with the consistency analysis of the algorithm, followed by the robustness analysis.

\paragraph*{Consistency}
Suppose that $T=\Tpred$. 
\subparagraph*{Case 1. $s > 1$.}
We will examine this first case in much more detail compared to the following cases.
Let $r := \floor{k - s}$. Observe that, by the choice of $k$ and $s$, we have $2 \leq r \leq k - 2$. By definition of $r$, we also have $\delta^{k - s - r} \in [1, \delta)$. Let us first consider the execution of the algorithm when $\alpha < \delta^{k - s - r}$.
For each iteration~$i = 0, \ldots, r$, (provided that the algorithm enters this iteration without terminating earlier) the algorithm appends $\bopt(\alpha \delta^i)$ since $\alpha \delta^i \leq \alpha \delta^r < \delta^{k - s}$.
When the algorithm enters iteration $(r + 1)$, which is still in the first phase because $r \leq k - 2$,
the algorithm proceeds to the second phase since $\alpha \delta^{r+1} \geq \delta^{r + 1} \geq \delta^{k - s}$.
Then it will append $\opt(\Tpred)$ during the second phase. Appending $\opt(\Tpred)$ by itself is sufficient to cover $T=\Tpred$ days, and the algorithm terminates.

On the other hand, let us now consider the execution of the algorithm when $\alpha \geq \delta^{k - s - r}$. In iteration~$i = 0, \ldots, r - 1$, the algorithm appends $\bopt(\alpha \delta^i)$ since $\alpha \delta^i < \delta^r \leq \delta^{k - s}$, unless the algorithm terminates earlier than that. When the algorithm enters iteration~$r$, since $\alpha \delta^r \geq \delta^{k - s}$, it proceeds to the second phase, appends $\opt(\Tpred)$, and terminate there.

To sum, the algorithm appends $\bopt(\alpha \delta^i)$ in iterations $0,\ldots,r-1$ unless it terminated earlier than the iteration; in iteration $r$, the algorithm may append $\bopt(\alpha \delta^i)$ only if $\alpha < \delta^{k - s - r}$; the algorithm leaves the first phase after iteration $r-1$ or $r$, so no other iterations of the first phase are entered; the algorithm may append $\opt(\Tpred)$ during the second phase; it never enters the third phase. Therefore, the total expected cost incurred by the algorithm is bounded from above by
\begin{align}
& \int_1^\delta \sum_{i = 0}^{r - 1} \alpha \delta^i f(\alpha) d\alpha + \int_1^{\delta^{k - s - r}} \alpha \delta^r f(\alpha) d\alpha + \delta^k 
 = \frac{\delta^r - 1}{\ln \delta} + \frac{\delta^{k - s} - \delta^r}{\ln \delta} + \delta^k \label{eq:rand-cons1}\\
& \leq \left(1 + \frac{\delta^{-s}}{\ln \delta} \right) \optval(\Tpred). \nonumber
\end{align}

\subparagraph*{Case 2. $0 \leq s \leq 1$.}
Note that $k \geq 2$ by the choice of $k$, and $\delta^{1-s} \in [1, \delta]$. Let us consider the execution of the algorithm when $\alpha < \delta^{1 - s}$.
For each iteration $i = 0, \ldots, k - 1$, the algorithm appends $\bopt(\alpha \delta^i)$ since $\alpha \delta^i < \delta^{k - s}$. The algorithm then proceeds to the third phase, appends $\bopt(\alpha \delta^k)$, and then terminates (unless it terminated even earlier).

Let us now consider the execution when $\alpha \geq \delta^{1 - s}$.
For each iteration $i = 0, \ldots, k - 2$, the algorithm appends $\bopt(\alpha \delta^i)$ since $\alpha \delta^i < \delta^{k - 1} \leq \delta^{k - s}$. In iteration~$(k - 1)$, since $\alpha \delta^{k - 1} \geq \delta^{k - s}$, the algorithm proceeds to the second phase, appends $\opt(\Tpred)$, and terminates (unless it terminated even earlier).

We can thus conclude that the algorithm in expectation incurs the cost of at most
\begin{align}
& \int_1^\delta \sum_{i = 0}^{k - 2} \alpha \delta^i f(\alpha) d\alpha + \int_1^{\delta^{1 - s}} (\alpha \delta^{k - 1} + \alpha \delta^k) f(\alpha) d\alpha +  \int_{\delta^{1 - s}}^{\delta} \delta^k f(\alpha) d\alpha  \nonumber \\
& = \frac{\delta^{k - 1} - 1}{\ln \delta} + \frac{(\delta^{k - 1} + \delta^k)(\delta^{1 - s} - 1)}{\ln \delta} + s\delta^k \label{eq:rand-cons2}\\
& \leq \left(\frac{\delta + 1}{\ln \delta} \delta^{-s} + s - \frac{1}{\ln \delta} \right) \optval(\Tpred). \nonumber
\end{align}

\paragraph*{Robustness}
Let $\optval(T) = \delta^{\istar}$ for some $\istar \in \mathbb{R}_{\geq 0}$ and let $r := \floor{k - s}$.
As we did in the consistency analysis, we will describe the execution of the algorithm as if it terminates only after appending a (sub)solution covering $T$ days or more, in favor of the simplicity of analysis.
Note that the algorithm may terminate earlier than that, but this still gives a valid upper bound on the algorithm's output cost.

\subparagraph*{Case 1. $s = 0$ or $\istar < r - 1$.}
We claim that, in this case, the algorithm never enters the second phase. If $s = 0$, note that $\alpha \delta^i < \delta^{k-s}$ always holds for every $i = 0, \ldots, k - 1$ since $\alpha < \delta$. If $\istar < r - 1$, observe that $\floor{\istar} + 1 \leq r - 1 \leq k - 1$, implying that iteration~$(\floor{\istar} + 1)$ is in the first phase.\footnote{When we say an iteration~$x$ is in the first phase, we are not implying that the particular iteration is actually entered by the algorithm at some point of its execution: we are simply stating that $x\in\{0,\ldots,k-1\}$. Recall that the iterations in the first phase are named $0,\ldots,k-1$.} For each iteration $i = 0, \ldots, \floor{\istar} + 1$, we have $\alpha \delta^i \leq \alpha \delta^{r - 1} < \delta^{k - s}$, and hence, the algorithm appends $\bopt(\alpha \delta^i)$ instead of proceeding to the second phase. Observe that, after iteration~$(\floor{\istar} + 1)$, the algorithm terminates since $\alpha \delta^{\floor{\istar} + 1} \geq  \delta^{\istar }$.

For each iteration $i = 0, \ldots, \floor{\istar}$, the algorithm appends $\bopt(\alpha \delta^i)$. In iteration~$\floor{\istar}$, if $\alpha \geq \delta^{\istar - \floor{\istar}}$, the algorithm  terminates since it appends $\bopt(\alpha \delta^{\floor{\istar}})$ with $\alpha\delta^{\floor{\istar}}\geq \delta^{\istar}$. On the other hand, if $\alpha < \delta^{\istar - \floor{\istar}}$, the algorithm enters the next iteration, appends $\bopt(\alpha \delta^{\floor{\istar} + 1})$ and terminates.
Therefore, the total expected cost is bounded by
\begin{align}
& \int_1^{\delta} \sum_{i = 0}^{\floor{\istar}} \alpha \delta^i f(\alpha) d\alpha + \int_1^{\delta^{\istar - \floor{\istar}}} \alpha \delta^{\floor{\istar} + 1} f(\alpha) d\alpha \label{eq:rand-rob0-1} \\
& = \frac{\delta^{\floor{\istar} + 1} - 1}{\ln \delta} + \frac{\delta^{\istar - \floor{\istar}} - 1}{\ln \delta} \cdot \delta^{\floor{\istar} + 1} \leq \frac{\delta}{\ln \delta} \optval(T) \leq \rho(\delta, s) \optval(T), \nonumber
\end{align}
where the last inequality holds since $e^z \geq e z$ for all $z$. (By choosing $z := \delta^{-s}$, $\frac{e^{{\delta}^{-s}}}{e{\delta}^{-s}}\ge 1$.) \newline

In what follows, let us assume that $s > 0$. Observe that $2 \leq r \leq k - 1$.

\subparagraph*{Case 2. $r - 1 \leq \istar < r$.}
Remark that $r \leq k - 1$. Let $m := \min(\istar + 1, k - s)$. Note that $m \geq r$ since $\istar \geq r - 1$ and $k - s \geq r$, and $m-r\leq \istar-r+1<1$. Observe that $\floor{\istar} = r - 1$.

Let us first consider the execution when $\alpha < \delta^{m - r}$. For each iteration $i = 0, \ldots, r$ of the first phase, the algorithm appends $\bopt(\alpha \delta^i)$ since $\alpha \delta^i < \delta^m \leq \delta^{k - s}$. The algorithm terminates after iteration~$r$ since $\istar < r$.

For $\delta^{m - r} \leq \alpha < \delta^{\istar - r + 1}$, observe that this case is nonempty only when $k - s < \istar + 1$ (and hence $m -r=k-s-r<\istar-r+1$). For each iteration $i = 0, \ldots, r - 1$ of the first phase, the algorithm appends $\bopt(\alpha \delta^i)$. On the other hand, when it enters iteration~$r$, it now proceeds to the second phase since $\alpha \delta^r \geq  \delta^m = \delta^{k - s}$. It then appends $\opt(\Tpred)$ and terminates.

Finally, if $\alpha \geq \delta^{\istar - r + 1}$, the algorithm appends $\bopt(\alpha \delta^i)$ in iterations $1,\ldots,r - 1$. Moreover, it terminates in iteration~$(r - 1)$ since $\alpha \delta^{r - 1} \geq \delta^{\istar}$.

We can thus derive that the total expected cost incurred by the algorithm is bounded from above by
\begin{align}
& \int_1^\delta \sum_{i = 0}^{r - 1} \alpha \delta^i f(\alpha) d\alpha + \int_1^{\delta^{m - r}} \alpha \delta^r f(\alpha) d\alpha + \int_{\delta^{m - r}}^{\delta^{\istar - r + 1}} \delta^k f(\alpha) d\alpha \nonumber \\
& = \frac{\delta^r - 1}{\ln \delta} + \frac{\delta^{m} - \delta^r}{\ln \delta} + (\istar + 1 - m) \delta^k \leq \frac{\delta^m}{\ln \delta} + (\istar + 1 - m) \delta^k. \label{eq:rand-rob2-1}
\end{align}
If $\istar + 1 \leq k - s$ (and hence $m = \istar + 1$), the above equation can further be bounded by $\rho(\delta, s) \cdot \optval(T)$ since $e^{z} \geq e z$ for all $z := \delta^{-s}$.

Now let us assume $\istar + 1 > k - s$. Note that the right-hand side of~\eqref{eq:rand-rob2-1} can be written as follows:
\[
\left( \frac{\delta^{k - s - \istar}}{\ln \delta} + (\istar + 1 - k + s) \delta^{k - \istar} \right) \optval(T).
\]
Let us substitute $z := k - s - \istar$. The following technical lemma then completes the proof of this case.

\begin{lemma} \label{lem:rand-rob2}
Given fixed $\delta\geq e$ and $s>0$, let $g(z) := \frac{\delta^z}{\ln \delta} + (1 - z) \delta^{z + s}$ be a function of $z$. We then have $g(z) \leq \rho(\delta, s)$ for every $z$.
\end{lemma}
\begin{proof}
From the derivative $g'(z) = \delta^z - \delta^{z + s} + (1 - z) \delta^{z + s} \ln \delta = \delta^z(1-\delta^s+\delta^s {\ln \delta}-z\delta^s {\ln \delta})$, we can see that the maximum of $g$ is attained at $z = z_0:=1 + \frac{\delta^{-s} - 1}{\ln \delta}$ with value $\frac{\delta}{e \ln \delta} \cdot \frac{e^{\delta^{-s}}}{\delta^{-s}} = \rho(\delta, s)$, completing the proof of the lemma. Note that we have $g'(z)\geq 0$ for all $z<z_0$ and $g'(z)\leq 0$ for all $z>z_0$ since $\delta^z>0$ and $z\mapsto 1-\delta^s+\delta^s {\ln \delta}-z\delta^s {\ln \delta}$ is a decreasing function of $z$.
\end{proof}

\subparagraph*{Case 3. $r \leq \istar < k - s$.}
Recall that $r := \floor{k - s}$ and hence $r \leq k - 1$. For each iteration $i = 0, 1, \ldots, r - 1$ of the first phase, the algorithm appends $\bopt(\alpha \delta^i)$ and enters the next iteration since $\alpha \delta^i < \delta^r \leq \delta^{\istar} < \delta^{k - s}$.

In iteration~$r$, let us first consider the execution when $\delta^{\istar - r} \leq \alpha < \delta^{k - s - r}$. The algorithm  appends $\bopt(\alpha \delta^r)$ and terminates after this iteration since $\delta^{\istar} \leq \alpha \delta^r < \delta^{k - s}$. When $\alpha \geq \delta^{k - s - r}$, it proceeds to the second phase after this iteration since $\alpha \delta^r \geq \delta^{k - s}$; the algorithm then appends $\opt(\Tpred)$ and terminates in the second phase.

Lastly when $\alpha < \delta^{\istar - r}$, the algorithm appends $\bopt(\alpha \delta^r)$ in iteration~$r$ since $\istar < k - s$. The behavior of the algorithm from this point differs depending on the value of $s$. If $s > 1$, we have $r \leq k - 2$, implying that the algorithm enters iteration~$(r + 1)$ which is still in the first phase. Observe that the algorithm then proceeds to the second phase without appending in this iteration since $\alpha \delta^{r + 1} \geq \delta^{k - s}$. The algorithm then appends $\opt(\Tpred)$ and terminates in the second phase. On the other hand, if $s \leq 1$, this implies $r = k - 1$, showing that iteration~$r$ is the last iteration of the first phase and therefore the algorithm directly proceeds to the third phase, iteration~$k$. In iteration~$k$, the algorithm appends $\bopt(\alpha \delta^k)$ and terminates since $k > k - s > \istar$.

Let us now bound the robustness. If $s > 1$, we have the following upper bound on the total expected cost:
\begin{align*}
& \int_1^\delta \sum_{i = 0}^{r - 1} \alpha \delta^i f(\alpha) d\alpha + \int_1^{\delta^{\istar - r}} (\alpha \delta^r + \delta^k) f(\alpha) d\alpha + \int_{\delta^{\istar - r}}^{\delta^{k - s - r}} \alpha \delta^r f(\alpha) d\alpha + \int_{\delta^{k - s -r}}^{\delta} \delta^k f(\alpha) d\alpha \\
& = \frac{\delta^r - 1}{\ln \delta} + \int_1^{\delta^{k - s - r}} \alpha \delta^r f(\alpha) d\alpha + \left( \int_1^{\delta^{\istar - r}} f(\alpha) d\alpha + \int_{\delta^{k - s - r}}^\delta f(\alpha) d\alpha \right) \delta^k \\
& = \frac{\delta^r - 1}{\ln \delta} + \frac{\delta^{k-s} - \delta^r}{\ln \delta} + (\istar + 1 - k + s) \delta^k \leq \left( \frac{\delta^{k - s - \istar}}{\ln \delta} + (1 - (k - s - \istar)) \delta^{k - \istar} \right) \optval(T)\\
&\leq \rho(\delta, s) \optval(T), 
\end{align*}
where the last inequality comes from Lemma~\ref{lem:rand-rob2} by letting $z := k - s - \istar$.

If $s \leq 1$, recall that $r=k-1$. We then have
\begin{align*}
& \int_1^\delta \sum_{i = 0}^{k - 2} \alpha \delta^i f(\alpha) d\alpha + \int_1^{\delta^{\istar - k+1}} (\alpha \delta^{k-1} + \alpha \delta^k) f(\alpha) d\alpha + \int_{\delta^{\istar - k+1}}^{\delta^{1 - s}} \alpha \delta^{k-1} f(\alpha) d\alpha + \int_{\delta^{1 - s}}^{\delta} \delta^k f(\alpha) d\alpha \\
& = \frac{\delta^{k - 1} - 1}{\ln \delta} + \int_1^{\delta^{1 - s}} \alpha \delta^{k - 1} f(\alpha) d\alpha + \int_1^{\delta^{\istar - k + 1}} \alpha \delta^k f(\alpha) d\alpha + \int_{\delta^{1 - s}}^\delta \delta^k f(\alpha) d\alpha \\
& = \frac{\delta^{k - 1} -1}{\ln \delta} + \frac{\delta^{k-s} - \delta^{k-1}}{\ln \delta} + \frac{\delta^{\istar+1}-\delta^k}{\ln \delta} +s \delta^k \leq \left( \frac{\delta}{\ln \delta} + \frac{\delta^{-s} - 1}{\ln \delta} \delta^{k - \istar} + s \delta^{k - \istar} \right) \optval(T) \\
&\leq \left( \frac{\delta^{1-s}}{\ln \delta} + s \delta \right) \optval(T), 
\end{align*}
where the last inequality holds from the fact that $\istar \geq r = k - 1$. 
We now claim 
\begin{equation}\label{eq:rand-rob3-finaleq}
\frac{\delta^{1 - s}}{\ln \delta} + s \delta \leq \rho(\delta, s)
\end{equation}
for all $\delta \geq e$ and $s > 0$, which would complete the proof.

\begin{lemma} \label{prop:rand-rob3-1}
For every $z > 0$, we have
$
g(z) := e^{z - 1} - z^2 + z \ln z \geq 0.
$
\end{lemma}
\begin{proof}
Remark that the derivative and the second derivative of $g$ is given as follows: $g'(z) := e^{z - 1} - 2z + \ln z + 1$ and $g''(z) := e^{z - 1} + 1/z - 2.$
Note that, for all $z > 0$, $g''(z) = e^{z - 1} + 1/z - 2 \geq z + 1/z - 2 \geq 0,$ implying that $g'$ is nondecreasing over $z > 0$. Observe that $g'(1) = 0$.
Hence, the minimum value of $g$ is attained at $z = 1$, where $g(1) = 0$.
\end{proof}
Recall that $\rho(\delta, s) = \frac{\delta}{e \ln \delta} \cdot \frac{e^{\delta^{-s}}}{\delta^{-s}}$. Dividing both sides of \eqref{eq:rand-rob3-finaleq} by $\frac{\delta}{\delta^{-s} \ln \delta}$ yields $\delta^{-2s} + s \delta^{-s} \ln \delta \leq e^{\delta^{-s} - 1}$. This inequality holds from Lemma~\ref{prop:rand-rob3-1} by letting $z := \delta^{-s}$.

\subparagraph*{Case 4. $k - s \leq  \istar < k$.}
In this case, we will re-use the argument from the consistency analysis. The only difference of the current case from the consistency analysis is that $T$ is not equal to $\Tpred$. In fact, we have $T<\Tpred$. However, the only place where the fact $T=\Tpred$ was used in the previous analysis is the observation that appending $\opt(\Tpred)$ or $\bopt(\alpha\delta^k)$ causes the algorithm to terminate. Since $T<\Tpred$, appending one of these two (sub)solutions causes the algorithm to terminate in this case, too, and the upper bounds \eqref{eq:rand-cons1} and \eqref{eq:rand-cons2} continue to hold.

If $s > 1$, \eqref{eq:rand-cons1} implies that the total expected cost incurred by the algorithm is at most
\[
\frac{\delta^r - 1}{\ln \delta} + \frac{\delta^{k - s} - \delta^r}{\ln \delta} + \delta^k \leq \left(\frac{1}{\ln \delta} + \delta^s \right) \delta^{k - s} \leq \rho(\delta, s) \optval(T),
\]
where the last inequality follows from $\istar \geq k - s$ and Lemma~\ref{lem:rand-rob4-1} below.

If $s \leq 1$, we have from \eqref{eq:rand-cons2} that the total expected cost is at most
\[\frac{\delta^{k - 1} - 1}{\ln \delta} + \frac{(\delta^{k - 1} + \delta^k)(\delta^{1 - s} - 1)}{\ln \delta} + s\delta^k  \leq \left( \frac{\delta + 1}{\ln \delta} + s \delta^{s} - \frac{\delta^s}{\ln \delta} \right) \delta^{k - s} \leq \rho(\delta, s) \optval(T),
\]
where the last inequality follows from $\istar \geq k - s$ and Lemma~\ref{lem:rand-rob4-2} below.

\begin{lemma} \label{lem:rand-rob4-1}
For any $\delta \geq e$ and $s\in\mathbb{R}$, we have
$
\frac{1}{\ln \delta} + \delta^s \leq \frac{\delta}{e \ln \delta} \cdot \frac{e^{\delta^{-s}}}{\delta^{-s}}.
$
\end{lemma}
\begin{proof}
By multiplying both sides by $e \delta^{-s} \ln \delta > 0$ and substituting $z := \delta^{-s}$, it suffices to show that
$
\delta e^z - e z \geq e \ln \delta.
$
By taking the partial derivative of the left-hand side with respect to $z$, we can infer that the left-hand side is minimized at $z = \ln (e / \delta) = 1 - \ln \delta$.
\end{proof}
\begin{lemma} \label{lem:rand-rob4-2}
For any $\delta \geq e$ and $0 \leq s \leq 1$, we have
$
\frac{\delta + 1}{\ln \delta} + s \delta^s - \frac{\delta^s}{\ln \delta} \leq \frac{\delta}{e \ln \delta} \cdot \frac{e^{\delta^{-s}}}{\delta^{-s}}
$
\end{lemma}
\begin{proof}
By multiplying both sides by $e \delta^{-s} \ln \delta > 0$ and substituting $z := \delta^{-s}$ (where $s = - \ln z / \ln \delta$), it is sufficient to prove that, for every $z \in [1/\delta, 1]$,
$
g(z) := \delta e^z - e (\delta + 1) z + e \ln z +e \geq 0.
$
Observe first that $g(1) = \delta e - e(\delta + 1) + e = 0$ and
\[
g\left( 1/\delta \right) = \delta e^{1/\delta} - e \left(1 + 1/\delta \right) + e \ln(1/\delta) + e = \delta e^{1 / \delta} - e/\delta + e \ln(1/\delta).
\]
Remark that, by Lemma~\ref{prop:rand-rob3-1} with $z := 1/\delta$, we have $\frac{1}{\delta e} g(1/\delta) \geq 0$.

Let us now consider the partial derivative of $g$ with respect to $z$:
$
\frac{\partial g}{\partial z}  = \delta e^z + \frac{e}{z} - e(\delta + 1).
$
Since $e^z$ and $e/z$ are both strictly convex over $z > 0$, we can see that $\partial g/ \partial z$ is also strictly convex over $z > 0$. Note also that
$
\frac{\partial g}{\partial z}\hspace{-.5ex}\bigm|_{z=1} = \delta e + e - e(\delta + 1) = 0.
$
We can thus conclude that $g$ has at most two solutions where one is $z = 1$. Moreover, as we have $g(1 / \delta) \geq 0$, we can see that $g(z) \geq 0$ for all $z \in [1/\delta, 1]$, completing the proof.
\end{proof}

\subparagraph*{Case 5. $\istar \ge k$.}
Recall that our assumption is that the algorithm terminates only after appending a (sub)solution covering $T$ days or more.
When $\istar >k$, the algorithm never appends such a solution during the first and second phases, and the algorithm does proceed to the third phase.
This may not be the case when $\istar =k$ for a technical reason, but for the analysis's sake, we will just assume that the algorithm always proceeds to the third phase without getting prematurely terminated. This may overestimate the cost incurred by the algorithm, but still gives a valid upper bound.

We will bound the expected cost incurred during the first and second phases, separately from the cost incurred during the third one. 
In fact, we will re-use \eqref{eq:rand-cons1} and \eqref{eq:rand-cons2} again, as we did in the previous case.
When $s>1$, we derived \eqref{eq:rand-cons1} based on the observation that the algorithm always proceeds to the second phase and terminates after this phase. Therefore, \eqref{eq:rand-cons1} can be used as is to bound the expected cost of the first two phases. On the other hand, when $s\leq 1$, the derivation of \eqref{eq:rand-cons2} was based on the observation that the algorithm terminates after either the second phase or the third phase. As such, we will slightly modify \eqref{eq:rand-cons2} to remove the contribution from the third phase: the expected cost incurred during the first two phases when $s\leq 1$ is at most
\begin{equation} \label{eq:rand-rob5-2}
\int_1^\delta \sum_{i = 0}^{k - 2} \alpha \delta^i f(\alpha) d\alpha + \int_1^{\delta^{1 - s}} \alpha \delta^{k - 1} f(\alpha) d\alpha +  \int_{\delta^{1 - s}}^{\delta} \delta^k f(\alpha) d\alpha \leq \frac{\delta^{k - s}}{\ln \delta} + s\delta^k .
\end{equation}

Now let us focus on the expected cost the algorithm incurs during the third phase. A similar argument to Case 1 can be applied here. Consider how the algorithm behaves when it enters iteration~$\floor{\istar}$. If $\alpha \delta^{\floor{\istar}} < \delta^{\istar}$ (or $\alpha < \delta^{\istar - \floor{\istar}}$), the algorithm further enters iteration~$(\floor{\istar}+1)$ and terminates after it. However, if $\alpha \delta^{\floor{\istar}} \geq \delta^{\istar}$ (or $\alpha \geq \delta^{\istar - \floor{\istar}}$), the algorithm terminates after iteration~$\floor{\istar}$. Since the algorithm appends $\bopt(\alpha \delta^i)$ for iteration $i$ in the third phase, the contribution of the third phase is bounded from above by
\begin{equation} \label{eq:rand-rob5-3}
\int_1^\delta \sum_{i = k}^{\floor{\istar}} \alpha \delta^i f(\alpha) d\alpha + \int_1^{\delta^{\istar - \floor{\istar}}} \alpha \delta^{\floor{\istar} + 1} f(\alpha) d\alpha = \frac{\delta^{\istar + 1} - \delta^k}{\ln \delta}.
\end{equation}

Let us combine these bounds. If $s > 1$, \eqref{eq:rand-cons1} and~\eqref{eq:rand-rob5-3} yield the following upper bound on the total expected cost:
\begin{align*}
&\frac{\delta^{k - s}}{\ln \delta} + \delta^k + \frac{\delta^{\istar + 1} - \delta^k}{\ln \delta} = \frac{\delta^{\istar + 1} + (\delta^{-s} + \ln \delta - 1) \delta^k}{\ln \delta} \leq \frac{\delta + \delta^{-s} + \ln \delta - 1}{\ln \delta} \optval(T)\\
&=\frac{\delta + \delta^{-s} + \min(s,1)\ln \delta - 1}{\ln \delta} \optval(T),
\end{align*}
where the inequality holds since $\delta \geq e$ and $\istar \geq k$, and the last equality follows from $s>1$. On the other hand, if $0 \leq s\leq 1$, \eqref{eq:rand-rob5-2} and~\eqref{eq:rand-rob5-3} yield the following bound:
\begin{align*}
&\frac{\delta^{k - s}}{\ln \delta} + s \delta^k + \frac{\delta^{\istar + 1} - \delta^k}{\ln \delta} = \frac{\delta^{\istar + 1} + (\delta^{-s} + s \ln \delta - 1) \delta^k}{\ln \delta} \leq \frac{\delta + \delta^{-s} + s \ln \delta - 1}{\ln\delta} \optval(T)\\
&=\frac{\delta + \delta^{-s} + \min(s,1) \ln \delta - 1}{\ln\delta} \optval(T),
\end{align*}
where the inequality holds since $\delta^{-s} + s \ln \delta - 1 = e^{-s\ln\delta} -(-s\ln\delta+1)\geq 0$ and $\istar \geq k$.

The following lemma completes the proof for this case.
\begin{lemma}
For every $\delta \geq e$ and $s \geq 0$, we have
$
\frac{\delta + \delta^{-s} + \min(s, 1) \ln \delta - 1}{\ln \delta} \leq \frac{\delta}{e \ln \delta} \cdot \frac{e^{\delta^{-s}}}{\delta^{-s}}.
$
\end{lemma}
\begin{proof}
Consider both sides of the inequality as a function of $s$ by treating $\delta$ as a fixed constant. It is then easy to see that the left-hand side is decreasing over $s\geq 1$. The right-hand side on the other hand is increasing over $s\geq 1$ since $x\mapsto \frac{e^x}{x}$ is a decreasing function of $x$ for $0<x<1$, and $s\mapsto \delta^{-s}$ is a decreasing function of $s$ for $s \geq 1$. Note that $\delta^{-s}\in(0,1)$ for all $s\geq 1$. Therefore, it suffices to prove the given inequality only for $0 \leq s\leq 1$. Under this condition, the inequality to prove can be rewritten as 
$
\frac{\delta + \delta^{-s} + s\ln \delta - 1}{\ln \delta} \leq \frac{\delta}{e \ln \delta} \cdot \frac{e^{\delta^{-s}}}{\delta^{-s}}
$
by removing the $\min$ operator.

By multiplying both sides with $e \delta^{-s} \ln \delta > 0$ and letting $z := \delta^{-s}$ (and therefore $s = - \ln z / \ln \delta$), we can rearrange this inequality as $
g(z) := \delta e^z + e z \ln z - e z^2 - e (\delta - 1) z \geq 0
$, which we need to show for all $z\in[1/\delta,1]$. We will show this inequality instead for all $z\in(0,1]$.

The first and second derivative of $g$, which we treat as a function of $z$, are:
$g'(z)  = \delta e^z + e \ln z - 2ez + (2 - \delta) e$ and $g''(z) = \delta e^z + e/z -2e$.
Observe that
$
g''(z) = \delta e^z + e/z -2e \geq e (e^z + 1/z - 2) \geq e(z + 1/z -2) \geq 0,
$
where the first inequality follows from $\delta \geq e$. This implies that $g'$ is nondecreasing over $(0,1]$. Note that $g'(1) = 0$, and hence $g'(z) \leq 0$ for $z\in(0,1]$. This shows that the minimum of $g$ is attained at $z = 1$. Observe that $g(1) = 0$.
\end{proof}

\subsection{Choice of Parameters and Comparison to Lower Bound}
\begin{figure}
\centering
\includegraphics[width=.5\textwidth]{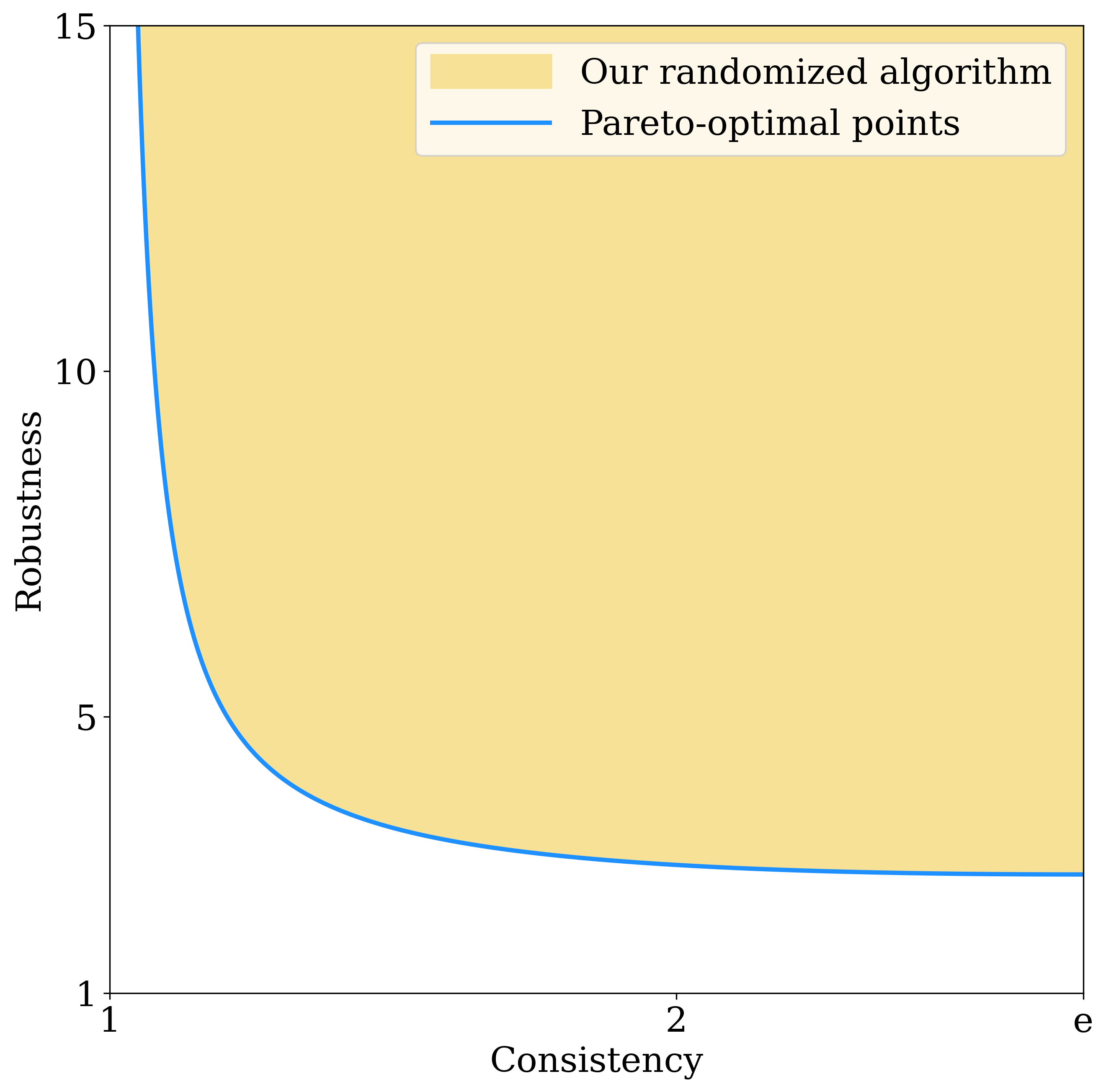}
\caption{The trade-off between consistency and robustness as $\delta$ and $s$ varies, shown as the yellow region. The blue solid line is the pareto-optimal points.}
\label{fig:con-tradeoff}
\end{figure}

Figure~\ref{fig:con-tradeoff} shows the trade-off between consistency and robustness offered by Theorem~\ref{thm:rand-param} as $\delta$ and $s$ varies. Each choice of the two parameters is shown as a point in the picture. Although these points form a  region in the graph, we would naturally want to  use only those choices of parameters that result in points on the boundary, shown as the blue solid line, which are pareto-optimal points.

We now compare the trade-off given by our algorithm against the lower bound presented in Section~\ref{sec:lb}.
To this end, we first obtain an alternative parametrization of the algorithm using a single parameter when the consistency is small.

\begin{theorem} \label{thm:rand-toff}
Let $\lambda^\star \approx 0.0861$ be the positive solution of $ \frac{\lambda + 1}{2} \cdot \ln \frac{2 \lambda}{\lambda + 1} = -1. $ Then, for $\lambda \in (0, \lambda^\star)$, there exists a randomized $(1 + \lambda)$-consistent $\frac{e(\lambda + 1)^2}{4\lambda}$-robust algorithm for the learning-augmented multi-option ski rental problem.
\end{theorem}
\begin{proof}
Let us choose $\delta := e^{2 / (\lambda + 1)}$ and $s := -\frac{\lambda + 1}{2} \cdot \ln \frac{2 \lambda}{\lambda + 1} > 1$. It is easy to verify that Theorem~\ref{thm:rand-param} gives $\chi(\delta, s) = 1 + \lambda$ and $\rho(\delta, s) = \frac{e(\lambda + 1)^2}{4\lambda}$.
\end{proof}
Note that, compared to the lower bound given by Theorem~\ref{thm:rand-buttonTradeoff}, the algorithm's robustness is within a factor of $e/2$.

\section{Lower Bound for Randomized Algorithms} \label{sec:lb}
In this section, we present the first nontrivial lower bound on the trade-off between consistency and robustness of randomized algorithms for the learning-augmented multi-option ski rental problem. The following theorem is to be shown.
\begin{theorem} \label{thm:rand-buttonTradeoff}
For all constant $\lambda \in (0,1)$ and $\varepsilon\in (0,1)$, any $(1+\lambda)$-consistent algorithm must have the robustness ratio greater than $\max\{\frac{(1+\lambda)^2}{2\lambda}, e\}-\varepsilon$.
\end{theorem}

The trivial bound of $e$ inherits from the lower bound on the competitive ratio (see Theorem 5 of \cite{shin2023improved}).
Therefore, it suffices to prove that any $(1+\lambda)$-consistent algorithm must have the robustness ratio greater than $\frac{(1+\lambda)^2}{2\lambda}-\varepsilon$.

Shin et al.~\cite{shin2023improved} consider the \emph{button problem} and give a linear program (LP) that yields a lower bound on the competitiveness of a randomized algorithm for this problem.
The button problem is defined as follows. We are given a list of $\nB$ buttons where each button $j$ is associated with a price $b_j$. The prices are monotone: $b_1 \leq \cdots \leq b_{\nB}$. Some buttons are designated as \emph{target} buttons, which form a suffix of the button list, i.e., there exists $\Jstar \leq J$ such that buttons $\Jstar$  through $\nB$ are all targets and none of the other buttons is a target. We can learn whether a button $j$ is a target or not only by pressing the button, at the price of $b_j$. We do not know ``the first target button'' $\Jstar$ but are given a prediction $\Jpred$ on $\Jstar$. The objective of this problem is to press one of the target buttons at the minimum total price.

This button problem is useful since the lower bound for this problem is (almost) inherited by the multi-option ski rental problem:
\begin{lemma}[\cite{shin2023improved}, Lemma 1] 
Suppose there exists a $\con$-consistent $\rob$-robust algorithm for the learning-augmented multi-option ski rental problem. Then there exists a $(\con+\varepsilon)$-consistent $(\rob+\varepsilon)$-robust algorithm for the button problem for all constant $\varepsilon\in (0,1)$.
\end{lemma}
Although any lower bound results on the button problem will immediately extend to the learning-augmented multi-option ski rental problem, Shin et al.~\cite{shin2023improved} unfortunately did not show any lower bounds on the consistency-robustness trade-off: they only showed a lower bound on the \emph{competitiveness} of randomized algorithms \emph{without} learning augmentation.

Before we prove Theorem~\ref{thm:rand-buttonTradeoff},
observe that
an algorithm's decision cannot be ``adaptive'' since the algorithm, until it presses a target and immediately terminates, will always learn that the button it pressed is not a target. As such,
any deterministic algorithm for the button problem is nothing more than a fixed sequence of buttons. The algorithm just presses the buttons according to this sequence until it eventually presses a target. We can assume without loss of generality that this sequence is increasing and the last button of the sequence is button~$\nB$, since the  target buttons form a suffix of the list.
A randomized algorithm can be viewed as a probability distribution over increasing sequence of buttons whose last button is button~$\nB$.

Let us consider the following instance of the button problem. The number of buttons $\nB$ will be chosen later as a sufficiently large number.
Let $b_j := j$ for every $j = 1, \ldots, \nB$. In what follows, we will always use $j$ itself instead of $b_j$ to denote the price of button $j$. The prediction given to the algorithm will always point to the last button $\nB$, i.e., $\Jpred = \nB$.
Note that we did not specify what the first target button $\Jstar$ is; in fact, we will consider a family of $J$ instances with $\Jstar = 1, \ldots, \nB$.

The following LP reveals a lower bound on the robustness of any $(1+\lambda)$-consistent randomized algorithm for this family of instances.
\begin{align*}
&\text{minimize} && \gamma &&\\
&\text{subject to} && \textstyle\sum_{j = 1}^\nB x_j = 1, && &&\\
& && \textstyle\sum_{j = t + 1}^{\nB} y_{t, j} = x_t + \sum_{j = 1}^{t - 1} y_{j, t}, && \forall t = 1, \ldots, \nB - 1 &&\\
& && \textstyle\sum_{j' = 1}^\nB {j'} \cdot \left(x_{j'} + \sum_{t = 1}^{\min(j, j') - 1} y_{t, {j'}} \right) \leq \gamma \cdot {j}, && \forall j = 1, \ldots, \nB, &&\\
& && \textstyle\sum_{j' = 1}^\nB {j'} \cdot \left(x_{j'} + \sum_{t = 1}^{\nB - 1} y_{t, {j'}} \right) \leq (1+\lambda) \cdot {\nB}, && &&\\
& && x_j \geq 0, && \forall j = 1, \ldots, \nB, &&\\
& && y_{t, j} \geq 0, && \begin{aligned} &\forall t = 1, \ldots, \nB - 1, \\ &\forall j = t + 1, \ldots, \nB. \end{aligned}&&
\end{align*}
In order to see that this indeed reveals a lower bound, fix an arbitrary $(1+\lambda)$-consistent randomized algorithm.
Let $x_j$ be the probability that button $j$ is the first button in the sequence, i.e., the first button pressed by the algorithm is button~$j$. For every $t$ and $j$ such that $t < j$, let $y_{t,j}$ be the  probability that buttons $t$ and $j$  appear consecutively in the sequence. In other words, $y_{t,j}$ is the marginal probability that the algorithm presses button $t$ immediately followed by button $j$, assuming that $t<\Jstar$.
We can now see that the first constraint requires that $\{x_j\}$ gives a probability distribution; the left-hand side and the right-hand side of the second set of constraints are two alternative ways of calculating the marginal probability that button $t$ appears in the sequence. The left-hand side of the third set of constraints is the expected cost of the algorithm's output when $\Jstar=j$, because $x_{j'} + \sum_{t = 1}^{\min(j, j') - 1} y_{t, {j'}}$ is the marginal probability that button~$j'$ is pressed when $\Jstar=j$. These constraints therefore ensure that $\gamma$ in an optimal solution is a lower bound on the robustness. The fourth constraint must be satisfied by (the probabilities exhibited by) any $(1+\lambda)$-consistent algorithm.

The dual of this LP is as follows.
\begin{align*}
& \text{maximize} && w - (1+\lambda)\nB\vhat && &&\\
& \text{subject to} && \textstyle\sum_{j = 1}^\nB j \cdot v_j = 1, && &&\\
& &&w \leq u_j + j \cdot \left( \vhat + \textstyle\sum_{j' = 1}^\nB v_{j'} \right), && \forall j = 1, \ldots, \nB - 1 &&\\
& &&w \leq {\nB} \cdot \left(\vhat + \textstyle\sum_{j' = 1}^\nB v_{j'} \right), && &&\label{prog:lp-dual-1}\tag{D1} \\
& &&u_t - u_j \leq j \cdot \left(\vhat + \textstyle\sum_{j' = t + 1}^\nB v_{j'} \right), && \begin{aligned} &\forall t = 1, \ldots, \nB - 2, \\ &\forall j = t + 1, \ldots, \nB - 1, \end{aligned} &&\\
& &&u_t \leq {\nB} \cdot \left(\vhat + \textstyle\sum_{j' = t + 1}^\nB v_{j'} \right), && \forall t = 1, \ldots, \nB - 1, &&\\
& &&w \in \mathbb{R}, && &&\\
& &&u_t \in \mathbb{R}, && \forall t = 1, \ldots, \nB - 1, &&\\
& &&v_j \geq 0, && \forall j = 1, \ldots, \nB, &&\\
& &&\vhat \geq 0.&& &&
\end{align*}

We will construct a feasible solution to this dual LP by constructing a solution to the following auxiliary LP first.
\begin{align*}
& \text{maximize} && w- (1+\lambda)\nB\vhat && &&\\
& \text{subject to} && w \leq u_j + j \cdot \left( \vhat + \textstyle\sum_{j' = 1}^\nB v_{j'} \right), && \forall j = 1, \ldots, \nB, &&\\
& && u_t - u_j \leq j \cdot \left(\vhat + \textstyle\sum_{j' = t + 1}^\nB v_{j'} \right), && \begin{aligned} &\forall t = 1, \ldots, \nB - 1, \\ &\forall j = t + 1, \ldots, \nB, \end{aligned} &&\label{prog:lp-dual-2}\tag{D2} \\
& && u_{\nB} = 0, && &&\\
& && w \in \mathbb{R}, && &&\\
& && u_t \in \mathbb{R}, && \forall t = 1, \ldots, \nB, &&\\
& && v_j \geq 0, && \forall j = 1, \ldots, \nB, &&\\
& && \vhat \geq 0. && &&
\end{align*}

Note that, as long as $\sum_{j=1}^\nB{j\cdot v_j}\neq 0$, any feasible solution to \eqref{prog:lp-dual-2} can be converted into a feasible solution to \eqref{prog:lp-dual-1} by dividing every variable by $\sum_{j=1}^\nB{j\cdot v_j}$.

Let us construct a solution to \eqref{prog:lp-dual-2}.
Let $\ell := \ceil{\frac{2\lambda}{1+\lambda}\nB}$. Note that $\ell \le \nB$ since $\lambda \in (0, 1)$. Let
\begin{align*}
v_j & := \begin{cases}1, &\text{ if } 1\le j \le \ell, \\ 0, & \text{ otherwise,} \end{cases}\\
\vhat &:= \nB-\ell, \\
u_t & := \nB(\nB-t), \text{ for all } t=1,\ldots,\nB \textrm{ and}  \\
w & := {\nB}^2.
\end{align*}
It is clear that the solution satisfies the last five sets of constraints.
The following two lemmas show that the above solution is indeed feasible to \eqref{prog:lp-dual-2}.
\begin{lemma}\label{lem:lb-feas-1}
For all $1\le t<j\le \nB$, $u_t - u_j \le j \cdot\left(\vhat + \sum_{j'=t+1}^{\nB}{v_{j'}}\right)$.
\end{lemma}
\begin{proof}
Remark that $\sum_{j'=t+1}^{\nB}{v_{j'}}=\ell-t$ if $t < \ell$, and $\sum_{j'=t+1}^{\nB}{v_{j'}}=0$ otherwise. We first bound from below the right-hand side by considering two cases. If $t < \ell$, then $j(\vhat + \ell-t) = j(\nB-t) = j\nB-jt \ge j\nB-\nB t = \nB(j-t)$; otherwise, $j\vhat = j(\nB-\ell) \ge j(\nB-t) \ge j\nB-\nB t = \nB(j-t)$. Combining with the fact that the left-hand side is equal to $\nB(j-t)$, the lemma follows.
\end{proof}

\begin{lemma} \label{lem:lb-feas-2}
For all $j = 1, \ldots, \nB$, $w \leq u_j + j \cdot \left( \vhat + \sum_{j' = 1}^{\nB} v_{j'} \right)$.
\end{lemma}
\begin{proof}
We have by construction $u_j + j \cdot \left( \vhat + \sum_{j' = 1}^{\nB} v_{j'} \right) = u_j + j (\vhat + \ell) = \nB(\nB-j)+j\nB=w$.
\end{proof}

We are now ready to prove Theorem~\ref{thm:rand-buttonTradeoff}.
Recall that we can construct a feasible solution to (\ref{prog:lp-dual-1}) by scaling down a feasible solution to (\ref{prog:lp-dual-2}). In light of this fact, it suffices to show that there always exists a family of instances such that $\displaystyle\frac{w- (1+\lambda)\nB\vhat}{\sum_{j=1}^\nB{j \cdot v_j}} \ge \frac{(1+\lambda)^2}{2\lambda}-\varepsilon$. Note that 
\begin{align}\label{eqn:lb-scaling}
\sum_{j=1}^\nB{j \cdot v_j} = \sum_{j=1}^\ell{j} & =\frac{\ell(\ell+1)}{2} \le \left(\frac{2\lambda}{1+\lambda}\nB+1\right)\left(\frac{\lambda}{1+\lambda}\nB+1\right)
\end{align} where the inequality follows from $\ell = \ceil{\frac{2\lambda}{1+\lambda}\nB} \le \frac{2\lambda}{1+\lambda}\nB+1$. We then have
\begin{align}
\frac{w- (1+\lambda)\nB\vhat}{\sum_{j=1}^\nB{j\cdot v_j}}
& = \frac{\nB^2-(1+\lambda)\nB(\nB-\ell)}{\sum_{j=1}^\nB{j \cdot v_j}} \nonumber \\
& \ge \frac{\nB^2-(1+\lambda)\nB\frac{1-\lambda}{1+\lambda}\nB}{\left(\frac{2\lambda}{1+\lambda}\nB+1\right)\left(\frac{\lambda}{1+\lambda}\nB+1\right)} \nonumber \\
& = \frac{\lambda \nB^2}{\left(\frac{2\lambda}{1+\lambda}\nB+1\right)\left(\frac{\lambda}{1+\lambda}\nB+1\right)}, \label{eqn:lb-ratio}
\end{align}
where the inequality follows from $\nB-\ell \le \nB\left(1-\frac{2\lambda}{1+\lambda}\right)=\frac{1-\lambda}{1+\lambda}\nB$ and \eqref{eqn:lb-scaling}.
By choosing  $\nB$ to be sufficiently large, we can see that (\ref{eqn:lb-ratio}) becomes arbitrarily close to $\frac{(1+\lambda)^2}{2\lambda}$. The conclusion follows from the weak LP duality.

\bibliographystyle{plain}
\bibliography{lit}

\end{document}